\documentclass[10pt,conference]{article}

\usepackage{spconf,amsmath,epsfig}
\usepackage{algorithm}
\usepackage{amsmath,amssymb,epsfig,graphics,subfigure}
\usepackage{theorem}
\usepackage{amssymb}

\DeclareMathSizes{5}{5}{3}{2}

\title{Robust Beamforming for Amplify-and-Forward MIMO Relay Systems Based on Quadratic Matrix Programming}
\name{Chengwen Xing, Shaodan Ma, Yik-Chung Wu and Tung-Sang Ng
 \thanks{This work was supported by the Hong Kong Research Grants Council (Grant No. 7154/08E).}}
\address{\\ Department of Electrical and Electronic Engineering
    \\The University of Hong Kong, Hong Kong
    \\ Email: \{cwxing, sdma, ycwu, tsng\}@eee.hku.hk}

\begin{document}
%
\maketitle

\begin{abstract}
In this paper, robust transceiver design based on
minimum-mean-square-error (MMSE) criterion for dual-hop
amplify-and-forward MIMO relay systems is investigated. The channel
estimation errors are modeled as Gaussian random variables, and then
the effect are incorporated into the robust transceiver based on the
Bayesian framework. An iterative algorithm is proposed to jointly
design the precoder at the source, the forward matrix at the relay
and the equalizer at the destination, and the joint design problem
can be efficiently solved by quadratic matrix programming (QMP).
\end{abstract}

%

\section{Introduction}
\label{sect:intro}

Recently, amplify-and-forward (AF) MIMO relay systems have gained
more and more attention from both academic and industrial
communities, due to its great potential to improve the wireless
channel reliability \cite{Chae08}, \cite{Stefania09} . For practical
applications, AF MIMO relay systems are to be adopted in future
communication protocols, such as Winner Project, LTE and
IMT-Advanced \cite{Stefania09}, to enhance the coverage of base
stations.

For transceiver design, joint LMMSE transceiver with perfect channel
state information (CSI) has been investigated in \cite{Rong09} and
an iterative algorithm has been proposed. Unfortunately, in
practice, CSI is generally obtained through estimation and perfect
CSI is very difficult to achieve. Robust transceiver design, which
could mitigate such performance degradation by taking the channel
estimation errors into account, is therefore of great importance and
highly desirable for practical applications.

In this paper, we consider robust linear transceiver design for AF
MIMO relay systems under imperfect CSI at both the relay and
destination. The precoder at the source, the forward matrix at the
relay and the equalizer at the destination are jointly designed
based on minimum-mean-square-error (MMSE)
criterion.\begin{figure}[!ht] \centering
\includegraphics[width=.45\textwidth]{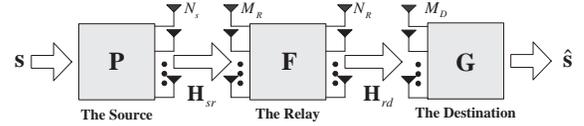}
\caption{Amplify-and-forward MIMO relay diagram.}\label{fig:1}
\end{figure} With the
channel estimation errors being modeled as Gaussian random
variables, robustness is incorporated into the optimization
objective function by taking expectation with respect to the channel
estimation errors. The joint design problem can efficiently solved
by quadratic matrix programming (QMP) \cite{Beck09}. Simulation
results show that the proposed robust algorithm performs better than
the transceiver design without taking channel estimation errors into
account.

The following notations are used throughout this paper. Boldface
lowercase letters denote vectors, while boldface uppercase letters
denote matrices. The notations ${\bf{Z}}^{\rm{T}}$,
${\bf{Z}}^{\rm{H}}$ and ${\bf{Z}}^*$ denote the transpose, Hermitian
and conjugate of the matrix ${\bf{Z}}$, respectively, and
${\rm{Tr}}({\bf{Z}})$ is the trace of the matrix ${\bf{Z}}$. The
symbol ${\bf{I}}_{M}$ denotes the $M \times M$ identity matrix,
while ${\bf{0}}_{M \times N}$ denotes the $M \times N$ all zero
matrix. The symbol ${\mathbb{E}}\{.\}$ represents the expectation
operation. The operation ${\rm{vec}}({\bf{Z}})$ stacks the columns
of the matrix ${\bf{Z}}$ into a single vector. The symbol $\otimes$
denotes the Kronecker product.

\section{System Model}

In this paper, a dual-hop amplify-and-forward (AF) cooperative
communication system is considered. In the considered system, there
is one source with $N_S$ antennas, one relay with $M_{R}$ receive
antennas and $N_R$ transmit antennas, and one destination with $M_D$
antennas, as shown in Fig.~\ref{fig:1}. At the first hop, the source
transmits data to the relay. The received signal, ${\bf{x}}$, at the
relay is
\begin{small}
\begin{align}
{\bf{x}}= {\bf{H}}_{sr}{\bf{P}}{\bf{s}}+{\bf{n}}_1
\end{align}\end{small}where ${\bf{s}}$ is the $N \times 1$ data vector transmitted by the
source with the covariance matrix
${\bf{R}}_{s}=\mathbb{E}\{{\bf{s}}{\bf{s}}^{\rm{H}}\}={\bf{I}}_N$,
${\bf{P}}$ is the precoder matrix with a transmit power constraint,
${\rm{Tr}}({\bf{P}}{\bf{P}}^{\rm{H}})\le P_s$, with $P_s$ is the
maximum transmit power at the source. The matrix ${\bf{H}}_{sr}$ is
the MIMO channel matrix between the source and the relay. Symbol
${\bf{n}}_1$ is the additive Gaussian noise with covariance matrix
${\bf{R}}_{n_1}$.

 At the relay, the received signal ${\bf{x}}$ is
multiplied by a forward matrix ${\bf{F}}$, under a power constraint
${\rm{Tr}}({\bf{F}}{\bf{R}_{x}}{\bf{F}}^{\rm{H}}) \le P_r$, where
${\bf{R}_{x}}=\mathbb{E}\{{\bf{x}}{\bf{x}}^{\rm{H}}\}$ and $P_r$ is
the maximum transmit power. Then the resultant signal is transmitted
to the destination. The received signal ${\bf{y}}$ at the
destination can be written as
\begin{small}
\begin{equation}
\label{equ:signal} {\bf{y}} = {{\bf{H}}_{rd} {\bf{F}}
{\bf{H}}_{sr}{\bf{P}}{\bf{s}}}  + {{\bf{H}}_{rd} {\bf{F}}{\bf{n}}_1
} + {\bf{n}}_2,
\end{equation}\end{small}
where ${\bf{H}}_{rd}$ is the MIMO channel matrix between the relay
and the destination, and ${\bf{n}}_2$ is the additive Gaussian noise
vector at the second hop with covariance matrix ${\bf{R}}_{n_2}$. In
order to guarantee the transmitted data ${\bf{s}}$ can be recovered
at the destination, it is assumed that $N_S$, $M_R$, $N_R$, and
$M_D$ are greater than or equal to $N$.

It is assumed that both the relay and destination have the estimated
channel state information (CSI). When channel estimation errors are
considered, we have $
 {\bf{H}}_{sr}={\bf{\bar H}}_{sr}+\Delta{\bf{H}}_{sr}$ and $
{\bf{H}}_{rd}={\bf{\bar H}}_{rd}+\Delta{\bf{H}}_{rd}$, where
${\bf{\bar H}}_{sr}$ and ${\bf{\bar H}}_{rd}$ are the estimated CSI,
while $\Delta{\bf{H}}_{sr}$ and $\Delta{\bf{H}}_{rd}$ are the
corresponding channel estimation errors whose elements are zero mean
Gaussian random variables. In general, the $M_R \times N_S$ matrix
$\Delta{\bf{H}}_{sr}$ can be written as
$\Delta{\bf{H}}_{sr}={\boldsymbol{\Sigma}}_{sr}^{\frac{1}{2}}{\bf{H}}_{W,sr}{\boldsymbol{\Psi}}_{sr}^{\frac{1}{2}}$
\cite{Ding09}, where the elements of the $M_R \times N_S$ matrix
${\bf{H}}_{W,sr}$ are independent and identically distributed
(i.i.d.) Gaussian random variables with zero mean and unit variance.
The $M_R \times M_R$ matrix ${\boldsymbol{\Sigma}}_{sr}$ and $N_S
\times N_S$ matrix ${\boldsymbol{\Psi}}_{sr}^{\rm{T}}$ are the row
and column covariance matrices of $\Delta{\bf{H}}_{sr}$,
respectively. The matrix $\Delta {\bf{H}}_{sr}$ is said to have a
matrix-variate complex Gaussian distribution, which can be written
as $\Delta {\bf{H}}_{sr}\sim \mathcal {C}\mathcal {N}_{M_R,N_S}
({\bf{0}}_{M_R \times N_S},{\boldsymbol{\Sigma}}_{sr} \otimes
{\boldsymbol{\Psi}}_{sr}^{\rm{T}} )$  \cite{Gupta00}. Similarly, for
the estimation error in the second hop, we have $ \Delta
{\bf{H}}_{rd} \sim \mathcal {C}\mathcal {N}_{M_D,N_R} ({\bf{0}}_{M_D
\times N_R} ,{\boldsymbol{\Sigma}} _{rd} \otimes {\boldsymbol{\Psi}}
_{rd}^{\rm{T}} )$, where the $M_D \times M_D$ matrix
${\boldsymbol{\Sigma}}_{rd}$ and $N_R \times N_R$ matrix
${\boldsymbol{\Psi}}_{rd}^{\rm{T}}$ are the row and column
covariance matrices of $\Delta{\bf{H}}_{rd}$, respectively. It is
assumed that ${\bf{H}}_{sr}$ and ${\bf{H}}_{rd}$ are estimated
independently, so the channel estimation errors, $\Delta
{\bf{H}}_{sr}$ and $\Delta {\bf{H}}_{rd}$, are independent.

\section{Problem Formulation}

At the destination, a linear equalizer ${\bf{G}}$ is adopted to
detect the transmitted data ${\bf{s}}$. The problem is how to design
the linear precoder matrix ${\bf{P}}$ at the source, the linear
forward matrix ${\bf{F}}$ at the relay and the linear equalizer
${\bf{G}}$ at the destination to minimize the mean square errors
(MSE) of the received data at the destination:
\begin{small}
\begin{align}
\label{equ:MSE_0_1} & {\rm{MSE}}({\bf{G}},{\bf{F}},{\bf{P}})
\nonumber\\
&
={\mathbb{E}}\{\|({\bf{G}}{\bf{H}}_{rd}{\bf{F}}{\bf{H}}_{sr}{\bf{P}}-{\bf{I}}_N){\bf{s}}
+{\bf{G}}{\bf{H}}_{rd}{\bf{F}}{\bf{n}}_1+{\bf{G}}{\bf{n}}_2 \|^2\}
\end{align}\end{small}where the expectation is taken with respect to ${\bf{s}}$, $\Delta
{\bf{H}}_{sr}$, $\Delta {\bf{H}}_{rd}$, ${\bf{n}}_1$ and
${\bf{n}}_2$. Since ${\bf{s}}$, ${\bf{n}}_1$ and ${\bf{n}}_2$ are
independent, the MSE expression (\ref{equ:MSE_0_1}) can be written
as
\begin{small}
\begin{align}
\label{MSE} &{\rm{MSE}}({\bf{G}},{\bf{F}},{\bf{P}}) \nonumber \\
&={\mathbb{E}}_{\Delta{\bf{ H}}_{rd},\Delta{\bf{
H}}_{sr}}\{{\rm{Tr}}(({\bf{ H}}_{rd}{\bf{F}}{\bf{
H}}_{sr}{\bf{P}})({\bf{ H}}_{rd}{\bf{F}}{\bf{
H}}_{sr}{\bf{P}})^{\rm{H}} )\}\nonumber \\
&\ \ +{\mathbb{E}}_{\Delta{\bf{ H}}_{rd}}\{{\rm{Tr}}(({\bf{G}}{\bf{
H}}_{rd}{\bf{F}}){\bf{R}}_{n_1}({\bf{G}}{\bf{
H}}_{rd}{\bf{F}})^{\rm{H}})\}+{\rm{Tr}}({\bf{G}}{\bf{R}}_{n_2}{\bf{G}}^{\rm{H}})
\nonumber \\
&\ \ +{\rm{Tr}}({\bf{I}}_N) -{\rm{Tr}}({\bf{G}}{\bf{\bar
H}}_{rd}{\bf{F}}{\bf{ \bar
H}}_{sr}{\bf{P}})-{\rm{Tr}}(({\bf{G}}{\bf{ \bar
H}}_{rd}{\bf{F}}{\bf{ \bar H}}_{sr}{\bf{P}})^{\rm{H}})
\end{align}\end{small}Because $\Delta{\bf{H}}_{sr}$ and $\Delta{\bf{H}}_{rd}$ are
independent, the first term of $\rm{MSE}$ is
\begin{small}
\begin{align}
\label{equ:MSE_1_a} &
{\mathbb{E}}_{\Delta{\bf{H}}_{sr},\Delta{\bf{H}}_{rd}} \{ {\rm{Tr}}
\left ( ({\bf{G}}{\bf{H}}_{rd}{\bf{F}}{\bf{H}}_{sr}{\bf{P}})
({\bf{G}}{\bf{H}}_{rd}{\bf{F}}{\bf{H}}_{sr}{\bf{P}})^{\rm{H}}
\right ) \} \nonumber \\
&= {\rm{Tr}}  ( {\bf{G}}{\mathbb{E}}_{{\Delta\bf{H}}_{rd}} \left\{
{\bf{H}}_{rd}{\bf{F}}{\mathbb{E}}_{{\Delta\bf{H}}_{sr}} \{
{\bf{H}}_{sr}{\bf{P}}{\bf{P}}^{\rm{H}}{\bf{H}}_{sr}^{\rm{H}}
\}{\bf{F}}^{\rm{H}} {\bf{H}}_{rd}^{\rm{H}}\right\} {\bf{G}}^{\rm{H}}
 ).
\end{align}\end{small}For the inner expectation, due to the fact that the distribution of
$\Delta{\bf{H}}_{sr}$ is matrix-variate complex Gaussian with zero
mean, the following equation holds \cite{Gupta00}
\begin{small}
\begin{align}
\label{equ:iteg} {\mathbb{E}}_{\Delta{\bf{ H}}_{rd}}\{{\bf{
H}}_{sr}{\bf{P}}{\bf{P}}^{\rm{H}}{\bf{ H}}_{sr}^{\rm{H}}\}
&={{\rm{Tr}}({\bf{P}}{\bf{P}}^{\rm{H}}{\boldsymbol
\Psi}_{sr}){\boldsymbol \Sigma}_{sr}+{\bf{\bar
H}}_{sr}{\bf{P}}{\bf{P}}^{\rm{H}}{\bf{\bar
H}}_{sr}^{\rm{H}}}\nonumber \\&\triangleq {\boldsymbol
\Pi}_{\bf{P}}.
\end{align}\end{small}Applying (\ref{equ:iteg}) and the corresponding result for $\Delta
{\bf{H}}_{rd}$ to (\ref{equ:MSE_1_a}), the first term of
${\rm{MSE}}$ in (\ref{MSE}) becomes
\begin{small}
\begin{align}
\label{equ:MSE_1} &{\rm{Tr}} \left ( {\bf{G}}{\mathbb{E}}_{\Delta
{\bf{H}}_{rd}} \left\{ {\bf{H}}_{rd}{\bf{F}}{\mathbb{E}}_{\Delta
{\bf{H}}_{sr}} \{
{\bf{H}}_{sr}{\bf{P}}{\bf{P}}^{\rm{H}}{\bf{H}}_{sr}^{\rm{H}}
\}{\bf{F}}^{\rm{H}}
{\bf{H}}_{rd}^{\rm{H}}\right\} {\bf{G}}^{\rm{H}} \right ) \nonumber \\
&={\rm{Tr}}( {\bf{G}}({\rm{Tr}}({\bf{F}}{\boldsymbol\Pi}_{\bf{P}}
{\bf{F}}^{\rm{H}} {\boldsymbol\Psi} _{rd}
   ){\boldsymbol\Sigma} _{rd}  + {\bf{\bar H}}_{rd}
   {\bf{F}}{\boldsymbol\Pi}_{\bf{P}}
   {\bf{F}}^{\rm{H}} {\bf{\bar H}}_{rd}^{\rm{H}}){\bf{G}}^{\rm{H}}
   ).
\end{align}\end{small}With similar calculations applied to the second term of MSE, the total MSE in
(\ref{MSE}) can be shown to be
\begin{small}
\begin{align}
\label{MSE_final} &{\rm{MSE}}({\bf{G}},{\bf{F}},{\bf{P}})={\rm{Tr}}
\left ( {\bf{G}}({\bf{\bar
H}}_{rd}{\bf{F}}{\bf{R}}_{\bf{x}}{\bf{F}}^{\rm{H}} {\bf{\bar
H}}_{rd}^{\rm{H}}+{\bf{K}}){\bf{G}}^{\rm{H}} \right )+{\rm{Tr}} (
{\bf{I}}_N)\nonumber
\\
&-{\rm{Tr}} \left ( {\bf{P}}^{\rm{H}}{\bf{\bar
H}}_{sr}^{\rm{H}}{\bf{F}}^{\rm{H}}{\bf{\bar
H}}_{rd}^{\rm{H}}{\bf{G}}^{\rm{H}} \right )-{\rm{Tr}} \left (
{\bf{G}}{\bf{\bar H}}_{rd}{\bf{F}}{\bf{\bar H}}_{sr}{\bf{P}}\right
),
\end{align}\end{small}where $
{\bf{R}}_{\bf{x}}={\boldsymbol\Pi}_{\bf{P}}+{\bf{R}}_{n_1}$ and
${\bf{K}}={\rm{Tr}}({\bf{F}}{\bf{R}}_{\bf{x}} {\bf{F}}^{\rm{H}}
{\boldsymbol\Psi} _{rd}
   ){\boldsymbol\Sigma} _{rd}+{\bf{R}}_{n_2}$. Notice that the matrix ${\bf{R}}_{\bf{x}}$ is the autocorrelation
matrix of the receive signal ${\bf{x}}$ at the relay. Finally, the
joint transceiver design can be formulated as the following
optimization problem
\begin{small}
\begin{align}
\label{MSE_op}
& {\min\limits_{{\bf{G}},{\bf{F}},{\bf{P}}}} \ \ \ {\rm{MSE}}({\bf{G}},{\bf{F}},{\bf{P}}) \nonumber \\
& {\rm{s.t.}} \ \ \ \ {\rm{Tr}}({\bf{P}}{\bf{P}}^{\rm{H}}) \le P_s,
\ \ \ {\rm{Tr}}({\bf{F}}{\bf{R}}_{\bf{x}}{\bf{F}}^{\rm{H}}) \le P_r.
\end{align}
\end{small}

\section{The Proposed Solution}

In this section, we derive an iterative algorithm to solve for
${\bf{P}}$, ${\bf{F}}$ and ${\bf{G}}$. In the following, it is shown
that given any two variables of ${\bf{P}}$, ${\bf{F}}$ and
${\bf{G}}$, the remaining one can be efficiently solved. Therefore,
the proposed algorithm computes ${\bf{P}}$, ${\bf{F}}$ and
${\bf{G}}$ iteratively, starting with initial values.


\underline{\textbf{Design of ${\bf{G}}$ :}}  When the precoder
${\bf{P}}$ at the source  and the forward matrix ${\bf{F}}$ at the
relay are fixed, the optimization problem (\ref{MSE_op}) is an
unconstrained convex optimization problem for ${\bf{G}}$. The
optimal equalizer ${\bf{G}}$ must satisfy ${\partial
{\rm{MSE}}({\bf{G}}},{{\bf{F}},{\bf{P}})}/{\partial {\bf{G}}^*} = 0$
which gives
\begin{small}
\begin{align}
\label{equ:G} {\bf{G}}= ({\bf{\bar H}}_{rd}{\bf{F}}{\bf{\bar
H}}_{sr}{\bf{P}})^{\rm{H}}({\bf{\bar
H}}_{rd}{\bf{F}}{\bf{R}}_{\bf{x}}{\bf{F}}^{\rm{H}} {\bf{\bar
H}}_{rd}^{\rm{H}}+{\bf{K}})^{-1}.
\end{align}
\end{small}

\underline{\textbf{Design of ${\bf{F}}$ :}} When ${\bf{P}}$ and
${\bf{G}}$ are fixed, the optimization problem (\ref{MSE_op})
becomes
\begin{align}
\label{MSE_F} & {\min}_{\bf{F}} \ \ \
{\rm{MSE}}({\bf{G}},{\bf{F}},{\bf{P}}) \ \ \  {\rm{s.t.}} \ \ \ \
{\rm{Tr}}({\bf{F}}{\bf{R}}_{\bf{x}}{\bf{F}}^{\rm{H}}) \le P_r.
\end{align}
Generally speaking, the optimization problem (\ref{MSE_F}) is a
quadratic matrix programming (QMP) problem with the variable
${\bf{F}}{\bf{R}}_{\bf{x}}^{1/2}$ and only one constraint. We can
formulate it into a semi-definite programming (SDP) problem to solve
for ${\bf{F}}$. However, because there is only one constraint, in
the following, we introduce another algorithm to compute ${\bf{F}}$
based on Karush-Kuhn-Tucker (KKT) conditions, and has a much lower
complexity. The corresponding KKT conditions of (\ref{MSE_F}) are
given as follows
\begin{small}
\begin{subequations}
\begin{align}
& {\bf{F}}=({\bf{\bar
H}}_{rd}^{\rm{H}}{\bf{G}}^{\rm{H}}{\bf{G}}{\bf{\bar
H}}_{rd}+{\boldsymbol \Psi}_{rd}{\rm{Tr}}({\bf{G}}{\boldsymbol
\Sigma}_{rd}{\bf{G}}^{\rm{H}})+\lambda{\bf{I}}_{N_R} )^{-1}\nonumber \\ & \ \ \ \ \ \
\ \ \ \ \ \ \ \ \ \ \ \ \ \ \ \ \ \ \ \ \ \ \ \ \ \  \  \ \times
{\bf{\bar H}}_{rd}^{\rm{H}}{\bf{G}}^{\rm{H}} {\bf{P}}^{\rm{H}}
{\bf{\bar
H}}_{sr}^{\rm{H}}{\bf{R}}_{\bf{x}}^{-1}. \label{KKT_1} \\
&
\lambda({\rm{Tr}}({\bf{F}}{\bf{R}}_{\bf{x}}{\bf{F}}^{\rm{H}})-P_r)=0
\label{KKT_2}\\
& \lambda \ge 0, \ \ \
{\rm{Tr}}({\bf{F}}{\bf{R}}_{\bf{x}}{\bf{F}}^{\rm{H}}) \le P_r
\label{KKT_3}.
\end{align}
\end{subequations}
\end{small}

Obviously from (\ref{KKT_1}), in order to compute the optimal
${\bf{F}}$, the Lagrangian multiplier $\lambda$ should be calculated
first. However, there is no closed-form solution of $\lambda$
simultaneously satisfying (\ref{KKT_2}) and (\ref{KKT_3}). Below we
propose a low complexity method to solve (\ref{KKT_2}) and
(\ref{KKT_3}). First, notice that in order to have (\ref{KKT_2})
satisfied, either $\lambda=0$ or
${\rm{Tr}}({\bf{F}}{\bf{R}_{x}}{\bf{F}}^{\rm{H}})=P_r$ must hold. If
$\lambda=0$ also makes (\ref{KKT_3}) satisfied, $\lambda=0$ is a
solution to (\ref{KKT_2}) and (\ref{KKT_3}). Since given ${\bf{G}}$
and ${\bf{P}}$, the optimization problem (\ref{MSE_F}) is a convex
quadratic programming problem of ${\bf{F}}$, which has only one
solution for ${\bf{F}}$, $\lambda=0$ is the only solution to
(\ref{KKT_2}) and (\ref{KKT_3}) in this case.

On other hand, if $\lambda=0$ does not make (\ref{KKT_3}) satisfied,
we have to solve
${\rm{Tr}}({\bf{F}}{\bf{R}_{x}}{\bf{F}}^{\rm{H}})=P_r$. It can be
proved that when ${\bf{G}}$ and ${\bf{P}}$ are fixed, the function
${\boldsymbol{f}}(\lambda)={\rm{Tr}}({\bf{F}}{\bf{R}}_{\bf{x}}{\bf{F}}^{\rm{H}})
$ is a decreasing function of ${\lambda}$ which satisfies
\begin{small}
\begin{align}
0\le {\lambda} \le \sqrt{{\rm{Tr}}({{\bf{\bar
H}}_{rd}^{\rm{H}}{\bf{G}}^{\rm{H}} {\bf{P}}^{\rm{H}} {\bf{\bar
H}}_{sr}^{\rm{H}}{\bf{R}}_{\bf{x}}^{-1}{\bf{\bar
H}}_{sr}{\bf{P}}{\bf{G}}{\bf{\bar H}}_{rd}})/{P_r}}.
\end{align}\end{small}Due to space limitation, the proof is not presented here.
Based on this result, $\lambda$ can be efficiently computed by a
one-dimension search, such as bisection search or golden search.
Since ${\rm{Tr}}({\bf{F}}{\bf{R}_{x}}{\bf{F}}^{\rm{H}})=P_r$ is a
stronger condition than
${\rm{Tr}}({\bf{F}}{\bf{R}_{x}}{\bf{F}}^{\rm{H}})\le P_r$,
(\ref{KKT_3}) is satisfied automatically in this case. In summary,
we take $\lambda=0$, if ${\boldsymbol f}(0) \le P_r$, and solve
${\boldsymbol f}(\lambda)=P_r$ otherwise.

\underline{\textbf{Design of ${\bf{P}}$ :}} When ${\bf{F}}$ and
${\bf{G}}$ are fixed, after a lengthy and tedious derivation, it can
be shown that the optimization problem (\ref{MSE_op}) is equivalent
to the following QMP problem \cite{Beck07}
\begin{small}
\begin{align}
\label{opt_P} & {\min\limits_{\bf{P}}} \ \ \
{\rm{Tr}}({\bf{P}}^{\rm{H}}{\bf{A}}_0{\bf{P}})+2{\mathcal{R}}({\rm{Tr}}({\bf{B}}_0^{\rm{H}}{\bf{P}}))
+{c_0} \nonumber \\
& {\rm{s.t.}} \ \ \ \
{\rm{Tr}}({\bf{P}}^{\rm{H}}{\bf{A}}_1{\bf{P}})+2{\mathcal{R}}({\rm{Tr}}({\bf{B}}_1^{\rm{H}}{\bf{P}}))
+{c_1} \le 0\nonumber \\
&\ \ \ \ \ \ \ \ \ \
{\rm{Tr}}({\bf{P}}^{\rm{H}}{\bf{A}}_2{\bf{P}})+2{\mathcal{R}}({\rm{Tr}}({\bf{B}}_2^{\rm{H}}{\bf{P}}))
+{c_2} \le 0,
\end{align}
\end{small} where the parameters are defined as follows
\begin{small}
\begin{align}
& {\bf{A}}_0={\boldsymbol \Psi}_{sr}{\rm{Tr}}({\bf{F}}{\boldsymbol
\Sigma}_{sr}{\bf{F}}^{\rm{H}}{\bf{M}})+{\bf{\bar H
}}_{sr}^{\rm{H}}{\bf{F}}^{\rm{H}}{\bf{M}}{\bf{F}}{\bf{\bar H}}_{sr},
\nonumber \\
& {\bf{M}}\triangleq {\boldsymbol
\Psi}_{rd}{\rm{Tr}}({\bf{G}}{\boldsymbol
\Sigma}_{rd}{\bf{G}}^{\rm{H}})+{\bf{\bar
H}}_{rd}^{\rm{H}}{\bf{G}}^{\rm{H}}{\bf{G}}{\bf{\bar H}}_{rd},
\nonumber \\
& {\bf{B}}_0=-({\bf{G}}{\bf{\bar H}}_{rd}{\bf{F}}{\bf{\bar
H}}_{sr})^{\rm{H}},
c_0={\rm{Tr}}({\bf{G}}({\bf{R}}_1+{\bf{R}}_{n_2}){\bf{G}}^{\rm{H}})+{\rm{Tr}}({\bf{I}}_{N}),
\nonumber\\ &{\bf{R}}_1\triangleq
{\rm{Tr}}({\bf{F}}{\bf{R}}_{\rm{n_1}}
{\bf{F}}^{\rm{H}}{\boldsymbol{\Psi}} _{rd}
){\boldsymbol{\Sigma}}_{rd} + {\bf{\bar H}}_{rd} {\bf{F}}
{\bf{R}}_{n_1} {\bf{F}}^{\rm{H}} {\bf{\bar
H}}_{rd}^{\rm{H}}, \nonumber \\
& {\bf{A}}_1={\bf{I}}_{N_S}, \ \ \ {\bf{B}}_1={\bf{0}}_{N_S,N}, \ \
\ {c_1}=-P_s,
\nonumber \\
& {\bf{A}}_2={\boldsymbol \Psi}_{sr}{\rm{Tr}}({\bf{F}}{\boldsymbol
\Sigma}_{sr}{\bf{F}}^{\rm{H}})+{\bf{\bar
H}}_{sr}^{\rm{H}}{\bf{F}}^{\rm{H}}{\bf{F}}{\bf{\bar H}}_{sr},
\nonumber \\
& {\bf{B}}_2={\bf{0}}_{N_S,N}, \ \ \
c_2={\rm{Tr}}({\bf{F}}{\bf{R}}_{n_1}{\bf{F}}^{\rm{H}})-P_r.
\end{align}
\end{small}
It is known that QMP problems can be transformed into semi-definite
programming (SDP) problems which can be efficiently solved by
interior point polynomial algorithms \cite{Beck07}. Based on the
properties of Kronecker product and the following definition
\begin{small}
\begin{align}
{\boldsymbol \Omega}_i \triangleq \left[ {\begin{array}{*{20}c}
  {\bf{I}}_{N} \otimes {\bf{A}}_i & {\rm{vec}}({{\bf{B}}_i)}  \\
   {\rm{vec}}^{\rm{H}}({\bf{B}}_i) & {c_i}   \\
\end{array}} \right], \ \ \ i=0,1,2,
\end{align}\end{small}the optimization problem (\ref{opt_P}) is equivalent to
\begin{small}
\begin{align}
\label{SDP}
& {\min\limits_{\bf{X}}} \ \ \ {\rm{Tr}}({\boldsymbol \Omega}_0{\bf{X}}) \nonumber \\
& {\rm{s.t.}} \ \ \ \ {\rm{Tr}}({\boldsymbol \Omega}_1{\bf{X}}) \le
0, \ \ \ {\rm{Tr}}({\boldsymbol \Omega}_2{\bf{X}}) \le 0
\nonumber \\
&\ \ \ \ \ \ \ \ \ \ {\bf{X}}=[{\rm{vec}}^{\rm{T}}({\bf{P}}) \
1]^{\rm{T}}[{\rm{vec}}^{\rm{H}}({\bf{P}}) \ 1]
\end{align}\end{small}If the constraint ${\rm{Rank}}({\bf{X}})=1$ is relaxed
(it is a well-known semi-definite relaxation (SDR) \cite{Beck09},
\cite{Ye03}), we have the following SDP relaxation problem
\begin{small}
\begin{align}
\label{SDR}
& {\min\limits_{\bf{Z}}} \ \ \ {\rm{Tr}}({\boldsymbol \Omega}_0{\bf{Z}}) \nonumber \\
& {\rm{s.t.}} \ \ \ \ {\rm{Tr}}({\boldsymbol \Omega}_1{\bf{Z}}) \le
0, \ \ \  {\rm{Tr}}({\boldsymbol \Omega}_2{\bf{Z}}) \le 0
\nonumber \\
&\ \ \ \ \ \ \ \ \ \  [{\bf{Z}}]_{NN_s+1,NN_s+1}=1, \ \ {\bf{Z}}
\succeq 0,
\end{align}\end{small}where ${\bf{Z}}$ is a Hermitian matrix. Because the QMP problem (\ref{opt_P}) is a convex quadratic programming problem, the relaxation gap of SDR is zero. In other words, the
optimization problems (\ref{SDP}) and (\ref{SDR}) have the same
optimal solution \cite{Beck07}, \cite{Beck06}.

\underline{\textbf{Summary and Convergence Analysis:}} Initialize
${\bf{P}}$ and ${\bf{F}}$ which satisfy
${\rm{Tr}}({\bf{P}}{\bf{P}}^{\rm{H}})=P_s$ and
${\rm{Tr}}({\bf{F}}{\bf{R}}_{\bf{x}}{\bf{F}}^{\rm{H}})=P_r$. For
simplicity, we can take ${\bf{P}}\propto{\bf{I}}$ and
${\bf{F}}\propto{\bf{I}}$. Then the proposed iterative algorithm
proceeds between (\ref{equ:G}), (\ref{KKT_1}) and (\ref{SDR}), until
$\|{\rm{MSE}}_i-{\rm{MSE}}_{i-1}\|\le t$ where ${\rm{MSE}}_i$ is the
MSE (\ref{MSE_final}) in the $i$th iteration, and $t$ is a
threshold. Since for any two of the ${\bf{P}}$, ${\bf{F}}$ and
${\bf{G}}$ fixed, the optimization problem (\ref{MSE_op}) is a
convex problem for the remaining variable, the proposed algorithm is
an alternative projection algorithm which is guaranteed to
converges.

\section{Simulation Results and Discussions}

In this section, we will investigate the performance of the proposed
algorithm and for the purpose of comparison, the algorithm based on
the estimated channel only (without taking the channel errors into
account) \cite{Rong09} is also simulated. In order to solve the SDP
problem, the matlab toolbox CVX is used \cite{Grant07}. In the
following, we consider an AF MIMO relay system where the source,
relay and destination are equipped with same number of antennas,
i.e., $N_S=M_R=N_R=M_D=4$. The estimated channels ${\bf{H}}_{sr}$
and ${\bf{H}}_{rd}$ are randomly generated as

\begin{tiny}
\begin{align}
& {\bf{\bar H}}_{sr}=\left[ {\begin{array}{*{20}c}
   {{1.02 + .82i} } & { - .01- 0.61i} & {.12-.26i} & {.02  + .64i}  \\
   {.08  + .90i} & {.70 - 1.22i} & {.06 +.19i} & {.46  + .62i}  \\
   {1.43  -  1.23i} & {.71  - .70i} & { - .23  + .81i} & {.03  +.25i}  \\
   {.43  - .71i} & {1.56  - .23i} & {.29  +  1.30i} & { -.63 +.73i}  \\
\end{array}} \right], \nonumber \\
& {\bf{\bar H}}_{rd}=\left[ {\begin{array}{*{20}c}
   {1.01  -  1.22i } & {.36  - .29i} & {.08  +  .50i} & { - .01  + .37i}  \\
   {.89  -  1.23i} & {1.05  -  .06i} & {.32  -  .21i} & {.45  + .73i}  \\
   {-.50  +  .23i} & {-.45  - .14i} & {-.55  +  .42i} & {1.01  +  .23i}  \\
   {-1.00  +  .38i} & {- .54  + .31i} & { - .00  + 0.62i} & {.82  +  1.32i}  \\
\end{array}}\right] \nonumber .
\end{align}
\end{tiny}Here the channel estimation algorithm in \cite{Ding09} is adopted, the
correlation matrices of channel estimation errors are in the form
\cite{Ding09}.
\begin{small}
\begin{align}
& {\boldsymbol \Psi}_{sr}={\bf{R}}_{T,sr}, \ \ {\boldsymbol
\Sigma}_{sr}={\sigma}_{e}^2({\bf{I}}_{M_R}+{\sigma}_{e}^2{\bf{R}}_{R,sr}^{-1})^{-1}, \nonumber \\
&{\boldsymbol \Psi}_{rd}={\bf{R}}_{T,rd}, \ \ {\boldsymbol
\Sigma}_{rd}={\sigma}_{e}^2({\bf{I}}_{M_D}+{\sigma}_{e}^2{\bf{R}}_{R,rd}^{-1})^{-1}\nonumber,
\end{align} \end{small}where $\sigma_{e}^2$ denotes the estimation error
variance \cite{Ding09}. The matrices ${\bf{R}}_{T,sr}$ and
${\bf{R}}_{R,sr}$ are the transmit and receive correlation matrices
in the first hop, respectively, and similar definitions apply to
${\bf{R}}_{T,rd}$ and ${\bf{R}}_{R,rd}$ for the second hop. The
widely used exponential model \cite{Ding09} is chosen for the
transmit and receive channel correlation matrices, i.e., $
 ({\bf{R}}_{T,rd})_{i,j}=({\bf{R}}_{T,sr})_{i,j}={\alpha}^{|i-j|}$,
 $({\bf{R}}_{R,rd})_{i,j}=({\bf{R}}_{R,sr})_{i,j}={\beta}^{|i-j|}$
where $\alpha$ and $\beta$ are the correlation coefficients.

We define the signal-to-noise ratio for the source-relay link
(${\rm{SNR}}_{sr}$) as
${\rm{E}}_s/{\rm{N}}_1=P_s/{\rm{Tr}({\bf{R}}}_{n_1})$, and is fixed
as ${\rm{E}}_s/{\rm{N}}_1=30{\rm{dB}}$. At the source, four
independent data streams are transmitted. For each data stream,
$10^5$ independent QPSK symbols are transmitted. The ${\rm{SNR}}$
for the relay-destination link (${\rm{SNR}}_{rd}$) is defined as
${\rm{E}}_r/{\rm{N}}_2=P_r/{\rm{Tr}({\bf{R}}}_{n_2})$. Each point in
the following figure is an average of 1000 independent realization
of estimation errors.

\begin{figure}[!ht]
\centering
\includegraphics[width=.4\textwidth]{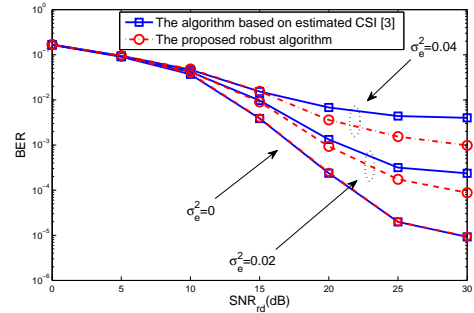}
\caption[scriptnote]{The BERs for the proposed iterative algorithm
and the algorithm based on estimated channels only for different
$\sigma_{e}^2$, when $\alpha=0.5$ and $\beta=0.4$.}\label{fig:2}
\end{figure}

Fig.~\ref{fig:2} shows the bit-error-rate (BER) performance of the
proposed algorithm and the algorithm based on estimated channels
only with different $\sigma_{e}^2$, when $\alpha=0.5$ and
$\beta=0.4$. It can be seen that when the channel estimation errors
decreases, the performances of both algorithms improve and they
coincide at $\sigma_e^2=0$. Furthermore, the performance of the
proposed algorithm is always better than that of the algorithm based
on estimated channels only.

\section{Conclusions}
In this paper, based on the Bayesian framework, robust linear
transceiver design for dual-hop AF MIMO relay systems has been
considered. The precoder matrix at the source, the linear forward
matrix at the relay and the linear equalizer at the destination have
been jointly designed based on minimum-mean-square-error (MMSE)
criterion. An iterative algorithm is proposed, and at each step, the
design problem can be formulated as a QMP problem which can be
efficiently solved. Simulation results showed that the performance
of the proposed robust algorithm is always better than that of the
algorithm based on estimated channels only.

\bibliographystyle{IEEEbib}
\bibliography[\small 

\end{document}